\documentclass[final,authoryear,5p,times,twocolumn]{elsarticle}
\usepackage{amsmath}
\usepackage{amssymb} 

\usepackage{tikz}
\usetikzlibrary{arrows,matrix,positioning,patterns}
\usetikzlibrary{calc}
\usepackage{pgfplots} 
\usetikzlibrary{external}
\usetikzlibrary{angles}
\usetikzlibrary{plotmarks}

\newdefinition{rmk}{Remark}
\newproof{pf}{Proof}
\newproof{pot}{Proof of Theorem \ref{thm2}}

\usepackage[ruled, vlined, linesnumbered]{algorithm2e}

\usepackage{amsthm}

\newtheorem{remark}{Remark}

\usepackage{amsmath}
\usepackage{amssymb} 

\usepackage{eurosym}
\usepackage{colortbl}

\usetikzlibrary{arrows.meta,calc,decorations.markings,math,arrows.meta}

\usepackage[labelformat=simple]{subcaption}
\DeclareCaptionLabelSeparator{periodspace}{.\quad}
\usepackage{color,framed}
\usepackage[utf8]{inputenc} 
\usepackage{tabularx}

\usepackage{rotating}
\usepackage{graphicx}
\usepackage{lscape}
\allowdisplaybreaks 
\usepackage{longtable}

\usepackage[utf8]{inputenc}
\usepackage{pgfplots}
\usepgfplotslibrary{groupplots} 
\pgfplotsset{compat=1.3}

\makeatletter
\def\ps@pprintTitle{%
 \let\@oddhead\@empty
 \let\@evenhead\@empty
 \def\@oddfoot{}%
 \let\@evenfoot\@oddfoot}
\makeatother

\begin{document}

\begin{frontmatter}
%
%
\title{Freeform Path Fitting for the Minimisation of the Number of Transitions\\ between Headland Path and Interior Lanes within Agricultural Fields}
\author{Mogens Graf Plessen\corref{cor1}}
\cortext[cor1]{\texttt{mgplessen@gmail.com}}

%
%
%
%

\begin{abstract}
Within the context of in-field path  planning this paper discusses freeform path fitting for the minimisation of the number of transitions between headland path and interior lanes within agricultural fields. This topic is motivated by two observations. Due to crossings of tyre traces such transitions in practice often cause an increase of compacted area. Furthermore, for very tight angles between headland path and interior lanes undesired hairpin turns may become necessary due to the limited agility of in-field operating tractors. By minimising the number of interior lanes both detrimental effects can be mitigated. The potential of minimising the number of interior lanes by freeform path fitting is evaluated on 10 non-convex real-world fields including obstacle areas, and compared to the more common technique of fitting straight interior lanes.
\end{abstract} 
\end{frontmatter}


\section{Introduction\label{sec_intro}}

\begin{figure*}
\centering
\includegraphics[width=7.8cm]{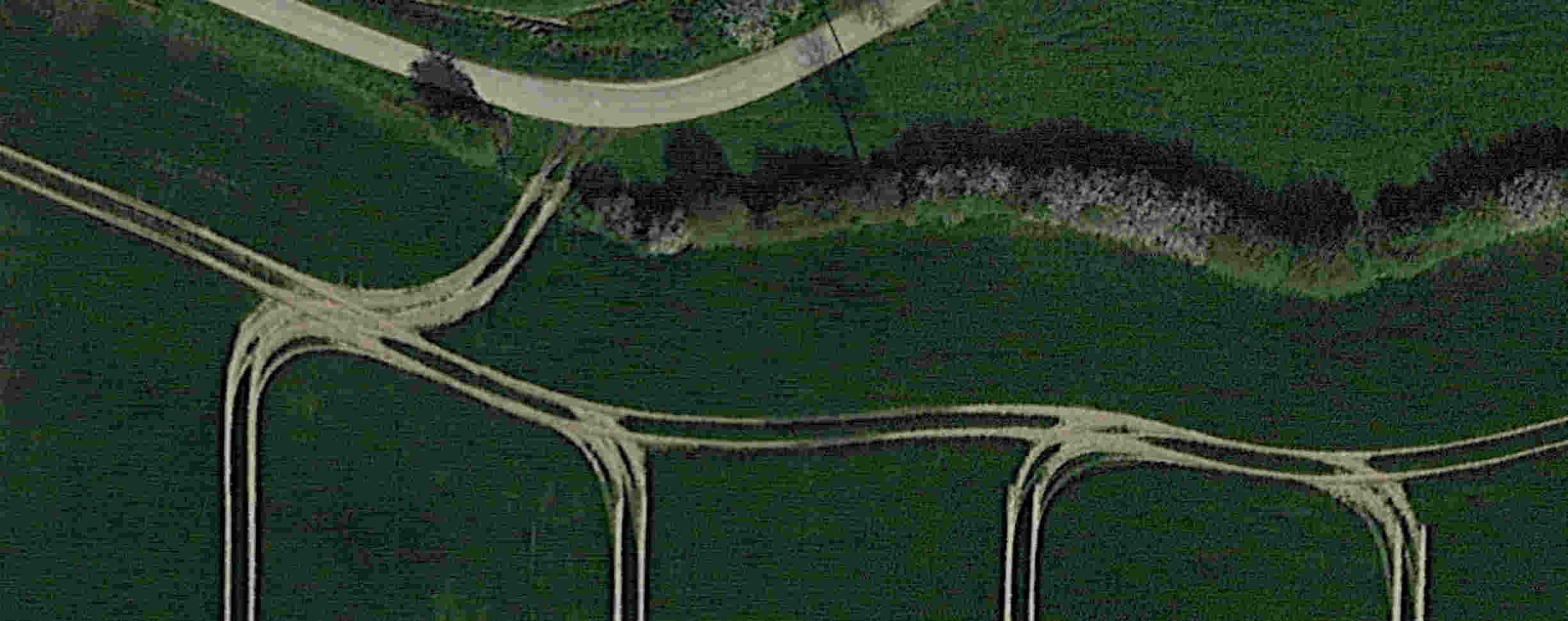}\hspace{1.5cm}
\begin{tikzpicture}
\draw [blue] plot [rounded corners=0.5cm] coordinates { (1,-0.1)(1,2)(3,2)(3,-0.1)};
\draw [blue] plot [rounded corners=0.5cm] coordinates { (-0.5,2)(4.5,2)};
%
%
\draw [draw=black,draw opacity=1, line width=10pt] plot [rounded corners=0.25cm] coordinates { (1,0.4)(1,0.8)};
%
\draw [draw=black,draw opacity=1, line width=1pt] plot [rounded corners=0.25cm] coordinates { (0,0.4)(2,0.4)};
\draw [draw=black,draw opacity=0.3, line width=26pt] plot [rounded corners=0.25cm] coordinates { (0,-0.1)(2,-0.1)};
\draw [black,-{Latex[scale=1.0]}] plot [rounded corners=0.25cm] coordinates { (1,0.8)(1,1.3)};
\node[color=black] (a) at (3.61, 1.01) {interior};
\node[color=black] (a) at (3.4, 0.66) {lane};
\node[color=black] (a) at (0.24, 0.81) {vehicle};
\node[color=black] (a) at (1.1, 2.3) {headland path};
\node[color=black] (a) at (1, -0.33) {Q};
\node[color=black] (a) at (3, -0.33) {R};
\node[color=black] (a) at (-0.8, 2) {C};
\node[color=black] (a) at (4.8, 2) {D};
\end{tikzpicture}
\caption{\emph{Left}: Visualisation of real-world transitions between headland path and interior lanes. As illustrated, the transitions between headland path and interior lanes can be ``messy'' in the sense that these often cause an increased amount of compacted field areas due to crossings of tyre traces. This motivates the minimisation of such transitions, and therefore to minimise the number of interior lanes as a proxy. \emph{Right}: Abstract visualisation with the definitions of headland path and interior lanes along which a vehicle (e.g., with spraying implement) might travel from location Q towards R or D. }
\label{fig_problFormulation}
\end{figure*}

\begin{figure}
\centering
\includegraphics[width=6cm]{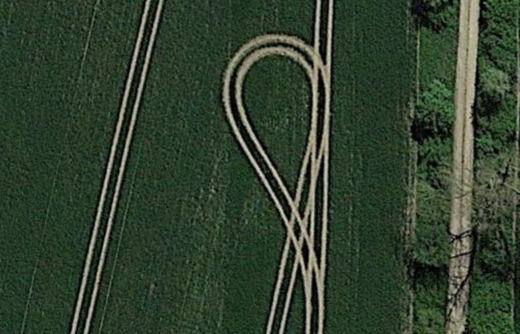}
\caption{For very tight angles between headland path and interior lanes undesired hairpin turns may become necessary due to the limited agility of in-field operating tractors. The more transitions between headland path and interior lanes the higher also the probability that such turns may occur, in particular, for complex fields. This provides the second motivation for the minimisation of the number of interior lanes by freeform path fitting.}
\label{fig_hairpin}
\end{figure}

\begin{table}
\centering
\begin{tabular}{|ll|}
\hline
\multicolumn{2}{|c|}{MAIN NOMENCLATURE}\\
\multicolumn{2}{|l|}{Symbols}\\
$d$ & Interpolation distance, (m).\\
$\epsilon$ & Hyperparameter for interpolation, (\%).\\
$N_l$ & Number of interior lanes, (-).\\
$\Delta N_l$ & Difference in number of interior lanes, (-) or (\%).\\
$\theta$ & Angle coordinate, ($^\circ$).\\
$\Delta \theta_\text{max}$ & Maximum permissible angle difference, ($^\circ$).\\
$w$ & Operating width (inter-lane distance), (m).\\
$(x,y)$ & Position coordinates, (m).\\[3pt]
\multicolumn{2}{|l|}{Abbreviations}\\
UTM & Universal Transverse Mercator coordinate system. \\
\hline
\end{tabular}
\end{table}

According to \cite{ahumada2009application} there are four main functional sectors for the agri-food supply chain: production, harvesting, storage and distribution. Optimising logistics and routing play an important role  in all of the four functional areas for improved supply chain efficiency. Furthermore, according to \cite{sorensen2010conceptual} it can be distinguished between in-field, inter-field, inter-sector and inter-regional logistics. This paper relates to the first functional area of the agri-food supply chain, i.e., production, and further to in-field logistics. The difficulty of in-field logistics arises from the vast variety of field shapes encountered in practice. In this perspective \cite{oksanen2013shape} presented eight indices for measuring the complexity of field shapes. For an application, see  \cite{janulevivcius2019estimation} who studied the effect of different field width-to-length ratios on tractor performance measures such as fuel consumption and exhaust emissions when ploughing. 

For in-field logistics it can be differentiated between three hierarchical planning layers: (i) the fitting of lanes within field contours, (ii) route planning for the traversal of these lanes, and (iii) trajectory planning accounting for agility and actuation constraints of the in-field operating vehicle to smooth out final paths. This paper focuses on the first hierachical layer and, in particular, how freeform path fitting compares to the more common method of fitting straights as interior lanes. For background, the coverage of agricultural fields growing, e.g.,  cereals or rapeseed requires lanes to be fitted within field contours such that in-field operating machinery can repeatedly travel along them during the work cycle after seeding and before harvest. These traversals typically occur many times throughout the year for multiple spraying and fertilizing applications.

\cite{oksanen2009coverage} presented two greedy algorithms for field
coverage path planning. The first algorithm splits a single field using a trapezoidal split-and-merge scheme into multiple smaller convex or near convex subfields that are then simpler to drive or operate using the best driving direction and best selection of subfields. Straight driving lanes are assumed. In contrast, in the second algorithm the path is planned on the basis of the machine’s current state and the search over a limited search horizon is on the next lanes instead of the next subfield. Lanes are now permitted to be curved. Note that for both algorithms multiple different path patterns may result within the same field. Thus, in general, also multiple new headland paths must be generated within the field in order to bound the different pattern regions, whereby segments of headland paths coincide for neighbouring regions. Consequently, more transitions between interior lanes and headlands and more compacted areas result. Thus, while partitioning of a field into multiple subfields may be inevitable for strongly non-convex field shapes that demand very different driving behaviour, it also comes at a cost. 

\cite{hameed2011driving} computed optimal driving directions for straight interior lanes based on the minimization of overlapped area using a genetic algorithm, before optimal routing is determined based on the minimization of the non‐working distance. Freeform path planning is here not discussed. Results are evaluated on two obstacle-free test fields.

\cite{hameed2010automated} presented a method for automated generation of guidance lines for operational field planning. A geometrical representation of the field is constructed as a geometrical entity
comprising discrete geometric primitives such as points, lines, and polygons. For their generation of parallel lanes the concept of ``longest edge'' of the field or partitioned subfields is crucial. It is selected to determine the driving direction. The curved edge is determined as a collection of sequential straight edges satisfying the criterion that the angle between two successive edges is less than or equal to a threshold. Complex field shapes are partitioned into a number of simpler subfields, typically convex polygons.

\cite{jin2011coverage} accounted for 3D terrain topography. Four critical tasks were therefore addressed: terrain modeling and representation, coverage cost analysis, terrain decomposition, and determining a suitable reference path such that by offsetting a lane pattern for full field coverage is obtained. Field boundary segments and topographic terrain contour lines were considered as the two categories for reference curve candidates. In \cite{guo2018application} a similar study is discussed, where soil and water conservation is also considered for the design of reference curves. In \cite{spekken2016planning} 3D terrain topograpgy is also examined. The objective was soil loss minimisation for sugarcane production. Therefore, ``hybrid'' curves were proposed as reference paths. These are generated by continuously offsetting multiple field contour segments towards each other until the hybrid curve is formed as their intersection points. While corresponding results were found to be beneficial, a disadvantage of this method is that the shape of the hybrid curve does not follow any particular segment of the field contour such that realisation in practice is difficult due to lack of nature-given visual reference landmarks.

Once lanes are fitted within field contours available methods from the literature can be employed for the two other aforementioned hierarchical planning layers (routing algorithms for traversal of lanes and trajectory planning for final path smoothing) including, for example, \cite{jensen2015coverage}, \cite{bochtis2013benefits}, \cite{hameed2016side}, \cite{yu2015optimization}, \cite{spekken2015cost}, \cite{seyyedhasani2018reducing}, \cite{plessen2018partial}, \cite{paraforos2018automatic} and \cite{backman2012path}.

To summarise, given UTM-coordinates of a field contour and of all its in-field obstacles, the first step is to fit interior lanes. Because of being the primary step lane fitting presents the foundation for all upstream logistical optimisation layers on top, including routing, trajectory planning and even multi-robot coordination. This underlines the importance of lane fitting. Furthermore, as satellite pictures show, the vast majority of fields in practice is fitted with straight lanes. Obviously, straights are attractive due to their simplicity and absence of turning. On the other hand, almost all fields are irregularly and very often even (strongly) non-convexly shaped. In view of this discrepancy the research question arises whether freeform path fitting may yet be underestimated for improved agricultural in-field operations and merit more studying. This argument is supported particularly  by the fact that straights can be considered as just a subset of the more general class of freeform paths. It is therefore expected that freeform paths can produce improved or at least equally good solutions for a variety of optimisation objectives. 

Within this context the motivation and contribution of this paper is to discuss the potential of freeform path fitting for the minimisation of the number of transitions between headland path and interior lanes within agricultural fields. As a proxy therefore, the number of interior lanes is minimised. This is achieved by optimised freeform fitting of interior lanes to arbitrarily non-convexly shaped field areas, that also may include obstacle areas. For two motivating visualisations see Figures \ref{fig_problFormulation} and \ref{fig_hairpin}. While freeform fitting of interior lanes is not new and sometimes in practice even performed intuitively by farmers, e.g., on wavy or curvedly shaped fields, the outcome of a quantitative comparison with respect to optimal straights fitting and explicit evaluation of the number of interior lanes that can be saved in real-world scenarios is not obvious.

The remaining paper is organised as follows: problem modeling and proposed solution, numerical results and the conclusion are described in Sections \ref{sec_solnMethod}-\ref{sec_conclusion}.

\section{Problem Modeling and Proposed Solution\label{sec_solnMethod}}

\newlength\figureheight
\newlength\figurewidth
\setlength\figureheight{8.3cm}
\setlength\figurewidth{8.3cm}
\begin{figure}
\centering
\input{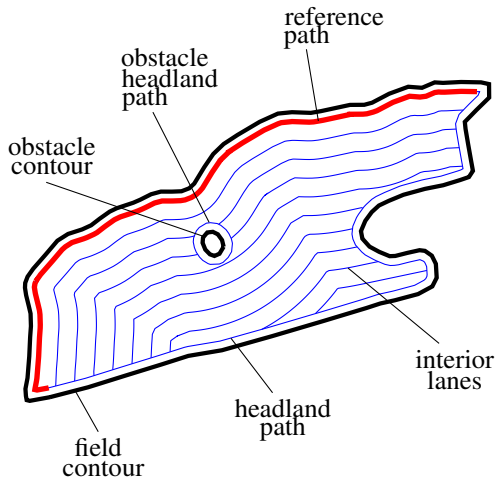}
\caption{Illustration of notation. Obstacles may represent tree islands, ponds, power pole masts, and so forth. The optimisation objective is to select the reference path (red) optimally as a partial segment of the headland path such as to minimise the total number of interior lanes subject to constraints.}
\label{fig_Field_sketch}
\end{figure}

Basic terminology is described in Figure \ref{fig_Field_sketch}. Problem input are location data of the \emph{field contour} and all available \emph{obstacle contours}. In a first step, the field contour is eroded to generate the \emph{headland path}. Similarly, obstacle contours are dilated to construct \emph{obstacle headland paths}. In a second step, a \emph{reference path} is selected as a partial segment of the headland path. In a third step, this reference is offset to generate a grid of \emph{interior lanes} fitted within field contours with inter-lane distance selected as the machinery operating width $w$. In this paper it is iterated over the selection of the reference path according to the criterion of minimising the total number of interior lanes denoted by $N_l$.

A candidate location point, $(x_k^{(i+1)},y_k^{(i+1)})$, in the next interior lane $i+1$ is generated based on two points, $(x_k^{(i)},y_k^{(i)})$ and $(x_{k+1}^{(i)},y_{k+1}^{(i)})$, in the previous interior lane $i$, according to
\begin{subequations}
\begin{align}
\theta_k^{(i)} &= \text{arctan}\left( \frac{y_{k+1}^{(i)} - y_{k}^{(i)}}{x_{k+1}^{(i)} - x_{k}^{(i)}} \right),\\
\begin{bmatrix} x_k^{(i+1)} \\ y_k^{(i+1)}  \end{bmatrix} &= \begin{bmatrix} \frac{x_k^{(i)} + x_{k+1}^{(i)}}{2} \\ \frac{y_k^{(i)} + y_{k+1}^{(i)}}{2} \end{bmatrix} + w \begin{bmatrix} \cos(\theta_k^{(i)} + \frac{\pi}{2}) \\ \sin(\theta_k^{(i)} + \frac{\pi}{2})  \end{bmatrix},
\end{align}
\label{eq_xykp1ip1}
\end{subequations}
before it is tested for pruning and constraints. Similarly the headland path and all obstacle headland paths are constructed, whereby the offsetting distance is here $w/2$, i.e., half the operating width. 

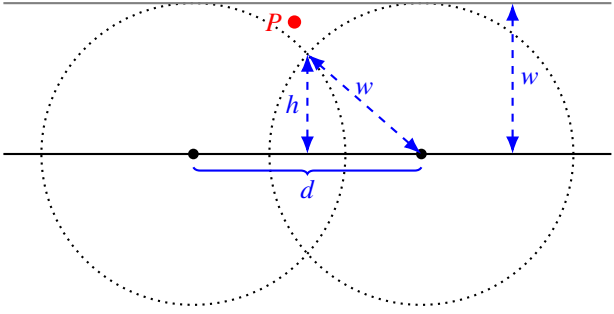
\begin{figure}
\centering
\vspace{0.cm}
\begin{tikzpicture}[thick,scale=1.0, every node/.style={scale=1.0}]
%
\draw[black] (-2.5,0) -- (5.5,0);
\draw[black,thick,dotted] (0,0) circle (2cm);
\draw[black,thick,dotted] (3,0) circle (2cm);
\fill[color=black] (0,0) circle [radius=2pt];
\fill[color=black] (3,0) circle [radius=2pt];
\draw[decoration={brace,mirror,raise=5pt},decorate,blue]
  (0,0) -- node[below=6pt] {$d$} (3,0);
%
\draw[black!50] (-2.5,2) -- (5.5,2);
\fill[color=red] (1.33,1.75) circle [radius=2.5pt];
\node[color=red] (a) at (1.05,1.75) {$P$};
%
\draw[blue,dashed,{Latex[scale=1.0]}-{Latex[scale=1.0]}] (1.5,0) --node[left=-0.7pt] {$h$} (1.5,1.3228756);
\draw[blue,dashed,{Latex[scale=1.0]}-{Latex[scale=1.0]}] (3,0) --node[above=0pt] {$w$} (1.5,1.3228756);
\draw[blue,dashed,{Latex[scale=1.0]}-{Latex[scale=1.0]}] (4.2,0) --node[right=-0.7pt] {$w$} (4.2,2);
\end{tikzpicture}
\caption{Derivation of spatial interpolation distance $d>0$.}
\label{fig_2circles}
\end{figure}

An implementation detail is discussed. After fitting of any interior lane its location data points are spatially extended by interpolation such that the distance between any 2 consecutive locations describing the interior lane is at most of length $d>0$. This is relevant for pruning and permits to work with \emph{circle inclusion checks} to determine if a candidate location $(x_{k}^{(i+1)},y_{k}^{(i+1)})$ according to \eqref{eq_xykp1ip1} maintains a distance of $w$ to all lane segments describing the previous interior lane $i$. Obviously this holds by definition for the lane segment described by $(x_{k}^{(i)},y_{k}^{(i)})$ and $(x_{k+1}^{(i)},y_{k+1}^{(i)})$. However, it may in general not hold for all lane segments and therefore may require pruning. The preferred method for the selection of $d>0$ is discussed next. See also Figure \ref{fig_2circles} for visualisation. The fundamental idea is to work with circle inclusion checks to determine if a candidate location point $P$ maintains with desired $\epsilon$-confidence (e.g., $\epsilon=99\%$) distance $w>0$ from any piecewise lane segment that concatenatedly describe the previous interior lane. With respect to Figure \ref{fig_2circles}, the constraint $h>\epsilon w$ can therefore be formulated. Using elementary geometric arguments this translates to a desired interpolation distance of 
\begin{equation}
d < 2 w \sqrt{1-\epsilon^2}.\label{eq_def_d}
\end{equation}
In contrast, for the headland and obstacle headland paths it is $d < 2 \frac{w}{2} \sqrt{1-\epsilon^2}$ because of their target-distance of half the operating width with respect to the field and obstacle contours, respectively. In final evaluation experiments it is set $w=36$m and $\epsilon=99$\%, which implies a spatial interpolation grid of at least 5m along headland and obstacle headland paths and 10m along interior lanes. Note that when collecting field and obstacle contour data points from real-world fields, these typically are already recorded to be sufficiently well shape-defining. Therefore, interpolation points along the field and obstacle contours are here added spatially only for improved constraint checking. The field and obstacle shape as defined by originally recorded contour data is not altered.

A pruned point does not abort interior lanes construction. It merely filters out some location points. In contrast, any constraint violation results in the dismissal of the entire corresponding reference path candidate. Before discussing constraints in detail, two more implementation details are therefore described. First, the last point of any interior lane is linearly extrapolated (leveraging the penultimate point for the direction) to obtain the intersection with the headland path. Second, after fitting of any interior lane its coordinates are extended by above discussed spatial interpolation technique with spacing according to \eqref{eq_def_d}.

Constraints are discussed. First, any crossings of any two interior lanes are prohibited. Second, too tight turns (expressed as the change of directions between two adjacent lane segments) are not admitted such that candidates with
\begin{equation}
|\theta_{k+1}^{(i)}-\theta_k^{(i)}|>\Delta \theta_\text{max},\label{eq_def_deltaThetaMax}
\end{equation}
for any $k=1,\dots,|\{\theta_k^{(i)}\}|-1,~\forall i=1,\dots,N_l$ are dismissed. In experiments the threshold was set to $\Delta \theta_\text{max}=135^\circ$. This somewhat aggressive choice is here motivated to determine a conservative upper bound on the saving potential of freeform path fitting. Third, also interpretable as a tight angle constraint to maintain a minimum $w$-distance among coordinates of each lane itself it is set 
\begin{equation}
\sqrt{(x_j^{(i)}-x_k^{(i)})^2 + (y_j^{(i)}-y_k^{(i)})^2}>w,
\end{equation}
for all $k=1,\dots,|\{x_k^{(i)}\}|,~\forall j=k+\Delta k,\dots,|\{x_k^{(i)}\}|,~\forall i=1,\dots,N_l$, and with hyperparameter $\Delta k>0$ denoting a blocking index interval. In experiments it was set $\Delta k=20$.

The operating width $w$ is given by available machinery hardware setup. For example, for spraying applications on fields growing cereals or rapeseed in central Europe it is often $w=36$m or $w=24$m. To sum up, for the proposed method three hyperparameters occur: $\epsilon\in[0,1]$, $\Delta \theta_\text{max}\in[0,\pi]$ and $\Delta k>0$. For the latter a lower bound can be determined as $\Delta k>\frac{w}{d}$. Then all three hyperparameters are lower bounded. The choice of $\epsilon=99\%$ as discussed above is reasonable for an accurate interpolation grid. To account for two operating widths one may heuristically select $\Delta k\approx \lceil\frac{2w}{d}\rceil$. Then, only one shaping hyperparameter, $\Delta \theta_\text{max}\in[0,\pi]$, remains.

As a proxy for the task of minimising the number of transitions between headland path and interior lanes the minimisation of the number of interior lanes, $N_l$, is selected as the optimisation criterion. This is valid since for every field run every interior lane must typically be covered exactly \emph{once}, thus requiring  exactly one entrance and one exit transition from and to the headland path, respectively.

Before discussing data-dependent results for 10 real-world fields, the comparative method of fitting \emph{straights} as interior lanes is reconsidered. 

\begin{remark}\label{def_remark1}
The method of fitting straights as interior lanes can, in general, be regarded as a special case of freeform path fitting. Instead of selecting a reference path defined by a segment or \emph{multiple} sequential location data points along the headland path for the latter scenario, it is defined by only \emph{two} points for the former case for the generation of straight interior lanes. For rectangular field shapes this method is optimal. However, for the general case of arbitrarily shaped fields optimality of straight interior lanes is not guaranteed. Since the problem class of straights fitting is included as a subset in the problem class of freeform path fitting, the latter will always be at least as good as the former for any objective, including, e.g., minimisation of total accumulated path length of interior lanes or total travel time along interior lanes. The critical disadvantage of freeform path fitting is the requirement of at least one shaping hyperparameter to constrain desired turning or curvatures of resulting paths. In contrast, for straights fitting no such shaping hyperparameter is required. Instead, it must solely be accounted for the physical machine operating width.
\end{remark}

\section{Numerical Results and Discussion\label{sec_IllustrativeEx}}

\setlength\figureheight{5cm}
\setlength\figurewidth{5cm}
\begin{figure*}
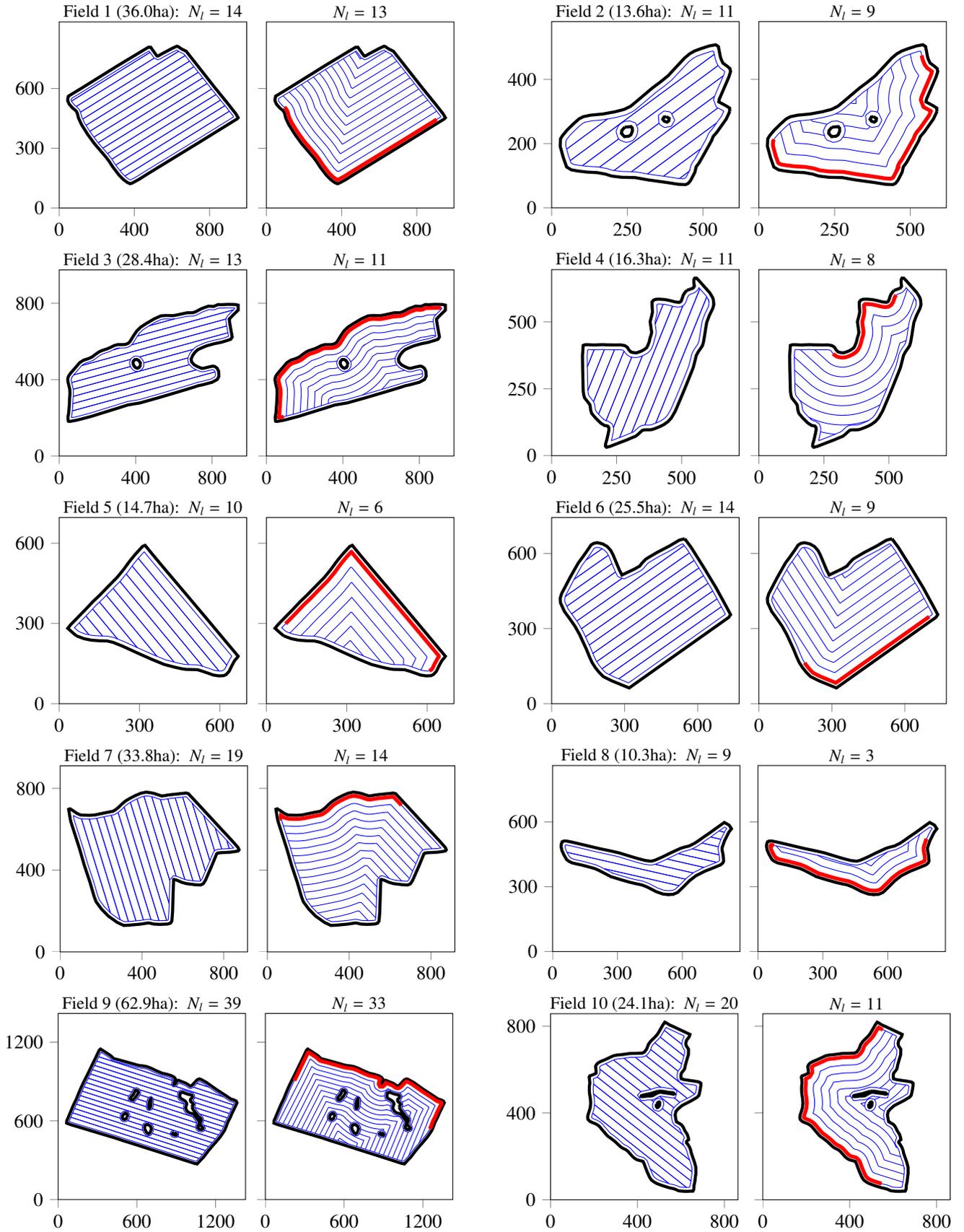

\centering
%
\input{Field1_NrLanes_STRAIGHTS.tex}\hspace{-0.4cm}\input{Field1_NrLanes_CURVED.tex}
~~~~~\input{Field2_NrLanes_STRAIGHTS.tex}\hspace{-0.4cm}\input{Field2_NrLanes_CURVED.tex}\\[-0.2cm]
\input{Field3_NrLanes_STRAIGHTS.tex}\hspace{-0.4cm}\input{Field3_NrLanes_CURVED.tex}
~~~~~\input{Field4_NrLanes_STRAIGHTS.tex}\hspace{-0.4cm}\input{Field4_NrLanes_CURVED.tex}\\[-0.2cm]
\input{Field5_NrLanes_STRAIGHTS.tex}\hspace{-0.4cm}\input{Field5_NrLanes_CURVED.tex}
~~~~~\input{Field6_NrLanes_STRAIGHTS.tex}\hspace{-0.4cm}\input{Field6_NrLanes_CURVED.tex}\\[-0.2cm]
\input{Field7_NrLanes_STRAIGHTS.tex}\hspace{-0.4cm}\input{Field7_NrLanes_CURVED.tex}
~~~~~\input{Field8_NrLanes_STRAIGHTS.tex}\hspace{-0.4cm}\input{Field8_NrLanes_CURVED.tex}\\[-0.2cm]
\input{Field9_NrLanes_STRAIGHTS.tex}\hspace{-0.4cm}\input{Field9_NrLanes_CURVED.tex}
~~~~~\input{Field10_NrLanes_STRAIGHTS.tex}\hspace{-0.4cm}\input{Field10_NrLanes_CURVED.tex}\\[-0.2cm]
\caption{Results for 10 real-world fields. The assumed operating width is $w=36$m. The optimal solutions of fitting straights and freeform paths as interior lanes for the minimisation of $N_l$  are displayed in the left and right subplot for each field, respectively. For color notation see Figure \ref{fig_Field_sketch}. Axes are denominated in meters.}
\label{fig_Fields}
\end{figure*}

\setlength\figureheight{4cm}
\setlength\figurewidth{9cm}
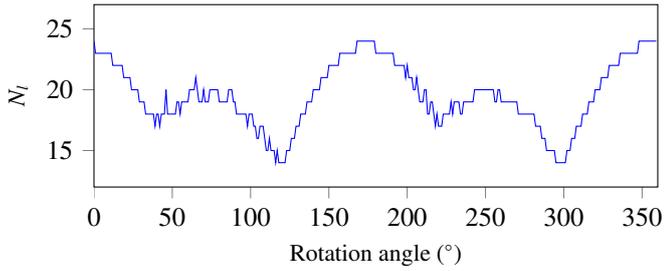
\begin{figure}
\centering
\begin{tikzpicture}

\begin{axis}[
xlabel={\small{Rotation angle ($^\circ$)}},
ylabel={\small{$N_l$}},
xmin=0, xmax=360,
ymin=12, ymax=27,
width=\figurewidth,
height=\figureheight,
tick align=outside,
tick pos=left,
x grid style={lightgray!92.02614379084967!black},
y grid style={lightgray!92.02614379084967!black}
]
\addplot [blue]
table {%
0 24
1 23
2 23
3 23
4 23
5 23
6 23
7 23
8 23
9 23
10 23
11 23
12 22
13 22
14 22
15 22
16 22
17 22
18 22
19 21
20 21
21 21
22 21
23 21
24 20
25 20
26 20
27 20
28 20
29 19
30 19
31 19
32 19
33 18
34 18
35 18
36 18
37 18
38 18
39 17
40 18
41 18
42 17
43 18
44 18
45 18
46 20
47 18
48 18
49 18
50 18
51 18
52 18
53 19
54 19
55 18
56 19
57 19
58 19
59 19
60 19
61 20
62 20
63 20
64 20
65 21
66 20
67 19
68 19
69 19
70 20
71 19
72 19
73 19
74 20
75 20
76 20
77 20
78 20
79 20
80 19
81 19
82 19
83 19
84 19
85 19
86 20
87 20
88 20
89 19
90 19
91 18
92 18
93 18
94 18
95 18
96 18
97 18
98 17
99 18
100 18
101 18
102 17
103 17
104 16
105 16
106 17
107 17
108 17
109 16
110 15
111 15
112 16
113 15
114 15
115 15
116 14
117 15
118 14
119 14
120 14
121 14
122 14
123 15
124 15
125 15
126 16
127 16
128 16
129 17
130 17
131 17
132 18
133 18
134 18
135 18
136 19
137 19
138 19
139 19
140 20
141 20
142 20
143 20
144 20
145 21
146 21
147 21
148 21
149 21
150 22
151 22
152 22
153 22
154 22
155 22
156 22
157 23
158 23
159 23
160 23
161 23
162 23
163 23
164 23
165 23
166 23
167 23
168 24
169 24
170 24
171 24
172 24
173 24
174 24
175 24
176 24
177 24
178 24
179 24
180 23
181 23
182 23
183 23
184 23
185 23
186 23
187 23
188 23
189 23
190 23
191 23
192 22
193 22
194 22
195 22
196 22
197 22
198 22
199 21
200 22
201 21
202 21
203 21
204 20
205 20
206 21
207 20
208 19
209 19
210 19
211 20
212 20
213 18
214 18
215 19
216 19
217 18
218 17
219 18
220 17
221 17
222 17
223 18
224 18
225 18
226 18
227 18
228 19
229 18
230 19
231 19
232 19
233 19
234 18
235 18
236 19
237 19
238 19
239 19
240 19
241 19
242 19
243 20
244 20
245 20
246 20
247 20
248 20
249 20
250 20
251 20
252 20
253 20
254 20
255 20
256 19
257 19
258 20
259 20
260 19
261 19
262 19
263 19
264 19
265 19
266 19
267 19
268 19
269 19
270 19
271 18
272 18
273 18
274 18
275 18
276 18
277 18
278 18
279 18
280 18
281 18
282 17
283 17
284 17
285 17
286 16
287 16
288 16
289 15
290 15
291 15
292 15
293 15
294 15
295 14
296 14
297 14
298 14
299 14
300 14
301 14
302 15
303 15
304 15
305 16
306 16
307 16
308 17
309 17
310 17
311 17
312 18
313 18
314 18
315 18
316 19
317 19
318 19
319 19
320 20
321 20
322 20
323 20
324 21
325 21
326 21
327 21
328 21
329 22
330 22
331 22
332 22
333 22
334 22
335 22
336 23
337 23
338 23
339 23
340 23
341 23
342 23
343 23
344 23
345 23
346 23
347 23
348 24
349 24
350 24
351 24
352 24
353 24
354 24
355 24
356 24
357 24
358 24
359 24
};

\end{axis}

\end{tikzpicture}
\vspace{-0.5cm}
\caption{Illustration of the effect of the rotation angle of straight interior lanes on the total number of interior lanes $N_l$ for Field 1 in Figure \ref{fig_Fields} and $w=36$m. Note that $N_l$ varies within a large range of 14 to 24 lanes.}
\label{fig_Field1_0to360_STRAIGHTS}
\end{figure}

\begin{table}
\vspace{0.3cm}
\centering
\begin{tabular}{|c|c|cc|cc|}
\hline
\rowcolor[gray]{1.0} Field & Size (ha) & $N_l^\text{straights}$ & $N_l^\text{freeform}$ & $\Delta N_l$ & $\Delta N_l$   \\[1pt] 
\hline
1 & 36.0 & 14 & 13 & -1 &  -7\% \\
2 & 13.6 & 11 & 9 & -2 &  -18\% \\
3 & 28.4 & 13 & 11 & -2 &  -15\% \\
4 & 16.3 & 11 & 8 & -3 &  -27\% \\
5 & 14.7 & 10 & 6 & -4 &  -40\% \\
6 & 25.5 & 14 & 9 & -5 &  -36\% \\
7 & 33.8 & 19 & 14 & -5 &  -26\% \\
8 & 10.3 & 9 & 3 & -6 &  -67\% \\
9 & 62.9 & 39 & 33 & -6 &  -15\% \\
10 & 24.1 & 20 & 11 & -9 &  -45\% \\
\hline
\end{tabular}
\caption{Summary of quantitative results displayed in Figure \ref{fig_Fields}.}
\label{tab_results}
\end{table}

The potential of minimising the number of interior lanes by freeform path fitting is evaluated on a variety of 10 real-world fields, and compared to the more common technique of fitting straight interior lanes. For the latter solution, the orientation of straights is also optimised to minimise the number of interior lanes $N_l$. Field sizes vary between 10.3ha and 62.9ha. Results are summarised in Figure \ref{fig_Fields} and Table \ref{tab_results}. The following observations are made.

First, savings range from -1 lane to -9 lanes or, perentage-wise speaking, the number of interior lanes could be reduced by -7\% to -67\% for freeform path fitting in comparison to optimal straights fitting. Note that for all fields a large operating width of $w=36$m is considered. For smaller operating widths savings scale linearly.

Second, while in some scenarios the optimised freeform path planning is intuitive such as for Field 8, it is surprising in other cases such as for Field 4. There are also subtle details about the exact optimal length of reference paths. In general, it therefore seems to be difficult to devise a reliable rule of thumb to select optimal reference paths. Thus, for general field shapes it clearly is best to turn to data-dependent numerical optimisation.

Third, the comparative method of fitting straights as interior lanes is discussed in more detail. For implementation simplicity it is in general desired to align straights to an approximately straight and typically also the longest segment of the field contour. For Field 1, Figure \ref{fig_Field1_0to360_STRAIGHTS} illustrates the resulting number of interior lanes $N_l$ as a function of the rotation angle of interior lanes, whereby 0$^\circ$ implies a vertical or y-axis aligned interior lane. The minimising solution is obtained for $N_l=14$ and is displayed in Figure \ref{fig_Fields}. When straight interior lanes were rotated by an additional $90^\circ$ to obtain alignment with the field contour in the north east, $N_l=17$ resulted. Then, savings for freeform path fitting would amount to -4 lanes and -24\%. This underlines the importance of careful selection of the orientation for straight interior lanes. To stress this more, when performing the grid search for Field 3 a range of $N_l\in[13,33]$ resulted, i.e., with possible variation of $\pm$20 interior lanes. Finally, a detail with respect to Figure \ref{fig_Field1_0to360_STRAIGHTS} is discussed. The lack of exact symmetry, for example, visible for rotation angles 40-120$^\circ$ and 220-300$^\circ$, is explained by the fact that the first interior lane is offset by distance $w$ from the headland path in the rotated coordinate system. Consequently, a residual distance results at the last othogonal interior lane with respect to the headland path. Since the field shape is not perfectly symmetric for Field 1, the resulting function in Figure \ref{fig_Field1_0to360_STRAIGHTS} is also not exactly repeating with a phase-shift of 180$^\circ$.

Fourth, the most obvious disadvantage of freeform path fitting is the increased amount of steering that is needed along interior lanes. This is largely relevant as long as there is a human vehicle driver who has to apply increased effort for lane tracking.

Fifth, as already pointed out in Remark \ref{def_remark1}, the strength of freeform path fitting is its flexibility in that it may (i) be used as a technique to also optimise \emph{alternative} objectives besides the number of interior lanes or to optimise a weighted trade-off among different criteria, and (ii) that solutions can easily be constrained to limit the desired amount of steering and to tailor results to the vehicle's agility capabilities. In particular, for $\Delta \theta_\text{max}=0$ in \eqref{eq_def_deltaThetaMax} the solution of fitting straights as interior lanes, i.e., the \emph{least-steering} solution, is recovered.

Ultimately, after having fitted interior lanes within field contours one can on top (i) determine field coverage path plans, e.g., according to the methods in \cite{plessen2018partial} and \cite{plessen2019optimal}, before (ii) smoothing out trajectories by accounting for actuation constraints of the in-field operating vehicle, for example, according to the control methods in \cite{backman2012navigation} or  \cite{plessen2017reference}.

\section{Conclusion\label{sec_conclusion}}

This paper contributed to the task of in-field path planning within agricultural fields by proposing a freeform path fitting method for the minimisation of the number of transitions between headland path and interior lanes. Therefore, as a proxy the minimisation of the number of interior lanes was selected as optimisation criterion. Spatial interpolation distances for pruning during the generation of interior lanes and constraints for the shaping of freeform paths were discussed. The potential of minimising the number of interior lanes by freeform path fitting was evaluated on 10 real-world fields and compared to the more common technique of fitting straight interior lanes. Field sizes varied between 10.3ha and 62.9ha with some including in-field obstacle areas. For an operating width of 36m optimal straights fitting resulted in a range of between 9 to 39 interior lanes. In comparison, freeform path fitting resulted in savings in the range of -1 lane to -9 lanes or, perentage-wise speaking, in a reduction of the number of interior lanes by -7\% to -67\%.

This paper focused on the very clear to quantify number of interior lanes as the optimisation criterion, which served as a proxy for the minimisation of the number of transitions between headland path and interior lanes. For future work an alternative objective such as the total accumulated path length along interior lanes may be considered. Then, ideally three consecutive layers are evaluated for determining the optimal grid of interior  lanes fitted within field contours. These three layers represent (i) selecting a reference path candidate as partial segment of the headland path and generating a corresponding grid of interior lanes as discussed in this paper, (ii) determining a routing solution for the coverage of all lane segments, before (iii) generating smoothed out trajectories accounting for agility and actuation constraints of the in-field operating vehicle, whereby the final detailed trajectory must also be planned such as to minimise spraying gaps. Similarly to as presented in this paper, it must then be iterated over reference path candidates before the total path length minimising solution is returned.

\bibliographystyle{model5-names} 
\bibliography{mybibfile.bib}
\nocite{*}







\end{document}